\documentclass[conference]{IEEEtran}

\usepackage[table]{xcolor}
\usepackage{tikz}
\usepackage{tcolorbox}
\usepackage{amsfonts}
\usepackage{subfig}
\usepackage{xurl}
\newcommand*\circled[1]{\tikz[baseline=(char.base)]{\node[shape=circle,fill,inner sep=0pt,minimum size=1pt] (char) {\textcolor{white}{#1}};}}
\usepackage[normalem]{ulem}

\begin{document}

\title{
Minimizing the Motion-to-Photon-delay (MPD) in Virtual Reality Systems}

\date{}

\author{\IEEEauthorblockN{ Akanksha Dixit} 
    \IEEEauthorblockA{\textit{Electrical Engineering} \\
    \textit{Indian Institute of Technology}\\
    New Delhi, India \\
    Akanksha.Dixit@ee.iitd.ac.in} 
    \and
    \IEEEauthorblockN{Smruti R. Sarangi} 
    \IEEEauthorblockA{\textit{Electrical Engineering} \\
    \textit{Indian Institute of Technology}\\
    New Delhi, India \\
    srsarangi@cse.iitd.ac.in}
   }

\maketitle

\thispagestyle{empty}

\begin{abstract}
	With the advent of low-power ultra-fast hardware and GPUs, virtual reality (VR) has gained a lot of prominence in the last few years and is being used in various areas such as education, entertainment, scientific visualization, and computer-aided design. VR-based applications are highly interactive, and one of the most important performance metrics for these applications is the motion-to-photon-delay (MPD). MPD is the delay from the user’s head movement to the time at which the image gets updated on the VR screen. Since the human visual system can even detect an error of a few pixels (very spatially sensitive), the MPD should be as small as possible.
	
	 Popular VR vendors use the GPU-accelerated Asynchronous Time Warp (ATW) algorithm to reduce the MPD. ATW reduces the MPD if and only if the warping operation finishes just before the display refreshes. However, due to the competition between applications for the shared GPU, the GPU-accelerated ATW algorithm suffers from an unpredictable ATW latency, making it challenging to find the ideal time instance for starting the time warp and ensuring that it completes with the least amount of lag relative to the screen refresh. Hence, the state-of-the-art is to use a separate hardware unit for the time warping operation.  Our approach, {\em PredATW}, uses an ML-based predictor to predict the ATW latency for a VR application, and then schedule it as late as possible. This is the first work to do so. Our predictor achieves an error of 0.77 ms across several  popular VR applications for predicting the ATW latency. As compared to the baseline architecture, we reduce deadline misses by 73.1\%. 
\end{abstract}

\section{Introduction}
\label{sec:Introduction}

Due to the introduction of numerous disruptive technologies over the past few years, computer graphics applications, such as VR \cite{9822241,9634089} applications have become incredibly popular.
Virtual Reality has revolutionized the gaming and entertainment sectors by allowing users to immerse themselves in a highly interactive environment \cite{Gupta2021How,THOMPSON2022VR}. It has also carved for itself an important place in medicine \cite{Virtual,Thomas2022Applications}. The global virtual reality market was valued at USD 21.83 billion in 2021 and is anticipated to expand at a CAGR of 15.0\% between 2022 and 2030 \cite{market}. Virtual reality-based applications are highly interactive and one of the most important performance metrics for these applications is the motion-to-photon-delay(MPD) \cite{Wagner2018MOTION, 2015Motion}. The MPD is the delay from the user’s head’s movement to the time at which the image gets updated on the VR screen. 
 
To get a good VR experience, the MPD should be as small as possible (elaborated in Section~\ref{sec:lowmpd}). Otherwise, users will perceive an anomalous view of the world, which may lead to  discomfort and visual disorientation. In the worst case, it may lead to more serious health consequences such as motion sickness and subsequent nausea \cite{ms1,ms2,ms3}. Many a time, the effects of a large and variable MPD are felt much later (we have personally experience this with Oculus Quest 2). Moreover, the applications where humans are at risk, for example, robotic surgery using VR, the MPD becomes more critical. Multiple VR devices such as Oculus Quest 2 \cite{oculus} and HTC Vive \cite{htc} have incorporated a technique broadly known as ATW (Asynchronous Time Warp) to decrease the MPD. Applications that use this technique are accelerated using GPUs; the overall aim is to realize a short MPD. 

\begin{figure*}[!htb]
	\centering
	\includegraphics[width=0.95\textwidth]{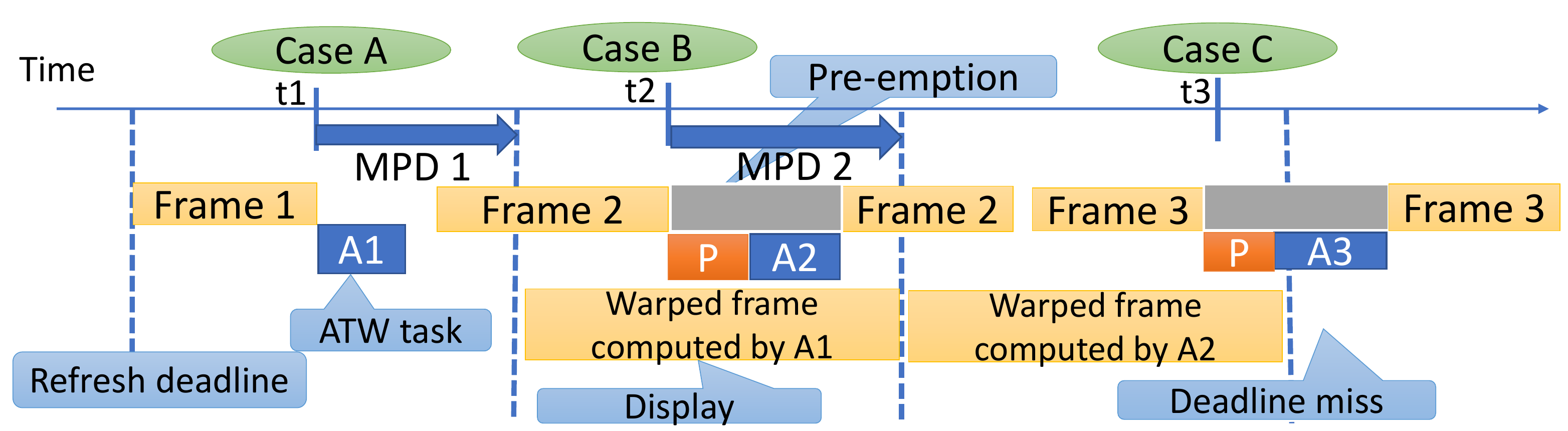}
	\caption{Possible scenarios in the GPU-accelerated ATW algorithm. $A1$, $A2$, and $A3$ are ATW tasks. $P$ is the pre-emption latency}
	\label{mpd}
\end{figure*}

Time warping is a type of image reprojection technique that maps the previously rendered frame to the correct position based on the latest head-orientation information and runs in parallel with the rendering process \cite{Waveren16}. It basically translates the image by certain pixels based on changes in the head position between the start of rendering and the initiation of the time warp \cite{2016Time} operation. The advantage of this is that the overall user experience is substantially better. 
Currently, VR headsets have a refresh rate of 90 Hz, which means  after every 11 ms a new frame is displayed on the screen. 
If rendering takes less than 11 ms, and we have adequate time left, then we can ``time warp'' the rendered image before the display
refresh deadline to make it more up to date. Whenever, rendering takes more time (exceeds 11 ms), we can ``time warp'' the 
previous frame and display it.

 This calls for an
asynchronous time warp (ATW) operation that can run concurrently \cite{2016Time}. Given that we cannot say for sure if a rendering task will complete by the deadline, at least finishing the ATW task is essential such that the last rendered frame can be transformed as per the current pose, which means that ATW needs to possibly preempt the rendering task and switch the context from rendering to warping. Therefore, the latency of the preemption and context switch is also added to the delay of ATW and this makes the ATW latency unpredictable because ATW uses the same graphics resources that the rendering process was using. Furthermore,  there could be other applications also running on the same GPU. This will further add to the noise. This is why current GPU-accelerated ATW solutions do not achieve the ideal MPD because of the unpredictable latency of the ATW operation~\cite{pim}.

One possible way to reduce the motion-to-photon-delay (MPD) significantly is to make the time warp operation fast. However, Xie et al. \cite{pim} show that applications can have a long MPD even with a fast implementation if the time warp is not initiated at the right time. They demonstrate that MPD is raised in both the situations: the time warp is completed before the display refreshes or if it misses the refresh deadline. Figure \ref{mpd} explains this. In the figure, we show three cases $A$, $B$, and $C$. In $A$, the time warping task ($A1$) finishes long before the refresh deadline and the warped frame that we see after the display refreshes corresponds to the head pose collected at time $t1$; this results in a long MPD. Hence, we should minimize the lag. For the cases $B$ and $C$, the rendering operation is interrupted to invoke ATW, but in the latter case, due to the long preemption latency $P$, the refresh deadline is missed and the user has a sub-optimal experience. So, the main problem here is to find the correct time instance to start the time warp operation. This is not straightforward because the ATW latency is variable.

VR applications are firm real-time systems, where deadlines are important most of the time (as we just saw). A lot of work has been done to decrease the MPD by offloading ATW from the GPU to a separate hardware component \cite{pim,HPTS21}. Some works focus on predictive tracking \cite{KunduRP21, HouZBD19} and motion prediction + a programmable display layer (PDL) \cite{SmitLBF09, SmitLF10}. 
\cite{PatneyKSKWBLL16, Luebke16} then try to optimize the rendering process itself to reduce the MPD. However, the problem of reducing the MPD when both ATW and the VR application share the same GPU is still \textit{open}. 

This forms the motivation for our work, {\em PredATW}. Our work aims to predict the time instance at which
the ATW operation should start so that a minimal MPD is obtained, and the frame rate and the quality are not affected adversely. To the best of our knowledge, we are the first to predict the ATW latency and run everything on the same GPU. We propose a simple decision tree-based predictor which uses frame-specific features (current frame or last rendered frame) such as the number of draw calls, brightness, and GPU execution time as the features for the predictor and the value of hardware performance counters.  The advantage of a decision-tree predictor is that our results are {\em explainable}.

Our primary contributions are:\\
\circled{1} We show that it is possible to predict the ATW latency on a GPU using simple metrics that are collected per frame and the values of performance counters. Furthermore, we show that the two most important metrics are the rendering time of the last rendered frame and the latency of the last ATW operation. \\
\circled{2} We compare the performance of various ML-based prediction models and show that a simple decision tree provides accurate results as compared to other ML-based models.\\
\circled{3} We were able to decrease the ATW refresh deadline miss rate by 73\% as compared to a GPU-accelerated ATW architecture. Our misprediction accuracy is a modest 0.77 ms in a frame size of 11.1 ms. 

The paper is organized as follows. Section \ref{sec:Background} provides the background of the VR architecture and ML-based models.  Section \ref{sec:Motivation} discusses the motivation for the work. The implementation details are given in Section \ref{sec:Implementation}. Section \ref{sec:Evaluation} shows the experimental results. We discuss related work in Section \ref{sec:RelatedWork} and finally we conclude in Section \ref{sec:Conclusion}.

\section{Background}
\label{sec:Background}
\subsection{The VR-Architecture Loop} 
\begin{figure}
\centering
 \includegraphics[width=0.48\textwidth]{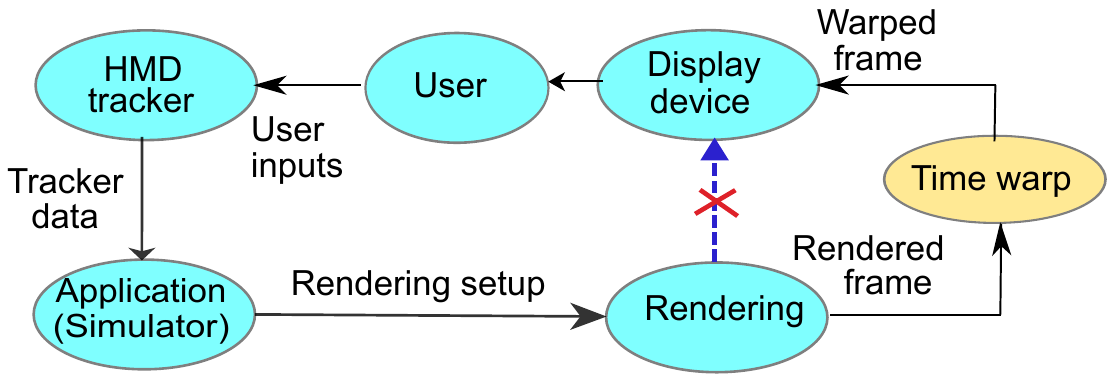}
\caption{Overview of the VR-architecture Loop}
\label{fig_1a}
\end{figure}

A brief overview of a typical VR-architecture loop is shown in Figure \ref{fig_1a}. In a VR system, the HMD tracker first gathers the user's position or motion data and then transmits that information to the application. The application copies the entire data to the GPU memory and sets up the rendering process. Finally, the rendering system generates two frames for the two eyes and displays those frames on the head-mounted displays (HMDs). In this process, a delay exists from the user’s head’s movement to the time at which the image gets updated on the VR screen; it is known as the motion-to-photon delay (MPD). The MPD should be as small as possible to provide a good user experience. Time warping is an image reprojection technique that is used to reduce this delay.

As discussed, time warping maps the previously rendered frame to the correct position based on the latest head-orientation information. Figure \ref{fig_1a} depicts a high-level overview of this warping architecture.  Note that ATW only applies to head rotational tracking, which means that the previously displayed frame is modified solely keeping in mind rotational changes of the head. 

\subsection{Inter-Frame Similarity}
Zhao et al. show \cite{ZhaoZBMYKSD20} that there exists Inter-Frame similarity in VR applications. We will focus on this aspect of VR applications to predict the ATW latency. This work uses two methods to show that inter-frame similarity exists in a VR application. The first method is to just compare the contents (after pre-processing). Whereas, the second method is to compare a few frame-specific features (image and runtime performance counter based). Let us discuss the first method here because that relies on prior work whereas the second method will be described in Section~\ref{sec:Motivation}, it requires some bespoke mechanisms that we develop.

\subsubsection{Method 1: Macroblock Similarity}
\label{2b}
To calculate the similarity of frame contents, we use the DCT (Discrete Cosine Transform) based approach mentioned in \cite{similar,histogram} to calculate the similarity between frames. 	In this approach, the computation of frame similarity is solely based on the DC (zero frequency) coefficients of the DCT since they represent a block's summary information. Since the human eyes are more sensitive to luminance than chrominance, the DC coefficients of only the brightness component are examined whereas the chrominance component is ignored. Once we know the DC coefficients of the two frames for which we want to compute the similarity, we analyze the similarity at the level of macroblocks to efficiently capture frame differences. A {\em macroblock} is a $2 \times 2$ matrix that has traditionally been used in linear block transform-based image and video compression \cite{Macroblock}. Let the two frames be $A$ and $B$. We construct macroblock matrices $M_1$...$M_n$ for each frame, where $n$ is the number of macroblocks (of DC coefficients). Two frames are considered to be similar if at least a minimum number of macroblocks are similar.

\subsection{ML-based Prediction Models}
\label{2c}
In machine learning, there are two types of prediction problems: classification and regression. In classification, we look for a model that can assist us with assigning a class to the datum. Regression, on the other hand, is a method for assigning continuous real values to  data rather than classes or labels. Since we are looking for the ATW latency, our problem falls under the regression category. A regression model uses the Mean Square Error (MSE) as the loss function. 

\subsubsection{Linear Regression: }
\label{2d}
Linear Regression is one of the simplest models used for regression. It tries to find a linear relationship between a dependent variable $Y$ and a set of independent variables $X$. In other words, it fits a straight line or a surface such that the errors between the actual values and the predicted values are minimized. 

\subsubsection{Decision Tree Regression: } 
The decision tree is used for both classification and regression tasks. One of the most essential aspects of this model is its explainability and the fact that its performance does not get affected by the non-linearity of the data. The decision tree builds models in the form of a tree. It splits the entire dataset into smaller subgroups in order to decrease the error. There are two sort of nodes in the tree: decision (internal) nodes and leaf nodes. The root node is a decision node at the top of the tree and covers the entire sample.  Each decision node has  a condition that determines which path to take, whereas the leaf node contains the final predicted outcome.

To build the model, we start with the root node. For deciding on a feature at the root node, we iterate through all the features and find the optimal split of the dataset. The splitting of the dataset happens until the error is minimized or a certain threshold is reached such as the a limit on the depth of the tree or the number of leaf nodes. The metric which is used to decide the splitting of the dataset is the sum of the squared error. This process is continued recursively. To predict the value for a given data point, we traverse through the entire tree following the conditions present at the decision nodes and reach a leaf node, which gives the final output.

\subsubsection{Random Forest Regression: } Random Forest, as the name implies, is a collection of decision trees. It combines the result of these decision trees to get the final result. This merging of the outcomes from various models or weak learners to give the final result refers to \textit{ensemble learning}. First, we randomly divide the dataset having $n$ points into $k$ parts. Then, decision trees are built independently for all the parts. During the prediction process, outputs of all those decision trees are averaged out to give the final outcome. The disadvantage of this model is that it needs rigorous training hence it is very slow and complex.  

\subsubsection{Gradient Boosting Regression: }
A gradient boosting regressor is also a variant of ensemble methods. It builds multiple regression trees and then combines their outcomes to produce final results. It differs from the random forest regressor in the way that it builds regression trees based on residuals; a residual is the difference between the actual output and the predicted output. These residuals and features are used to train regression trees and the residual predicted by these models are incorporated into the input model. This process is repeated several times and the input model is pushed towards the correct prediction.

\subsubsection{Convolutional Neural Network (CNN): }
A CNN is a type of neural network that is used for prediction purposes. We do not use CNNs in our case because of the following reasons.

\begin{itemize}
\item The structure of a CNN is very complex, each of its layers uses multiple multipliers and adders. Hence, it becomes slow and also requires a lot of storage. Since we want to predict the ATW latency so that the MPD is reduced, it is not suitable to have a predictor with a large delay.

\item The second reason is that CNNs are not explainable. We cannot tell how each feature is contributing to the final prediction. As we want to analyze the application's and the system's behavior, we need an explainable model.
\end{itemize} 
\section{Motivation}
\label{sec:Motivation}
In this section, first, we show the list of benchmarks that we use for experiments. Then, we evaluate the latency of time warping in a shared GPU environment. After that, we show that in a VR application, there exists a substantial similarity between consecutive frames, which our predictor needs to exploit. Finally, we note that since the resolution of the foveal part of the human eye is very high, there is a very strong need to reduce the MPD as much as possible -- this motivates our approach.

\subsection{Overview of the Benchmarks}
Table \ref{table_1} shows the benchmarks used in this study. Orange Room is a benchmark test included in the VRMark benchmark suite~\cite{vrmark}.
The rest of the benchmarks are well-known VR applications and taken from Steam similar to prior works~\cite{pim, HPTS21}. For a more accurate evaluation of the impact of the workload size on our solution, we render the applications at various resolutions (mentioned in the table).

\begin{table}[!htb]

\begin{center}
	
	\resizebox{0.99\columnwidth}{!}{
\begin{tabular}{ |l|l|l|l|l|l| }

\hline
\rowcolor[HTML]{EFEFEF}
\textbf{Abbr.} &
\textbf{Name} &
\textbf{Platform} &
\textbf{Engine} &
\textbf{API} &  
\textbf{Resolution} \\ 
\hline\hline
\em{IM} & InMind VR \cite{InMind} & Steam &  Unity & DX11 & 1920$\times$1080 \\
\em{IM2} & InMind VR 2&  &  & & 1290$\times$800 \\
\hline
\em{SH} & Shooter & Custom &  Unity & DX11 & 1920$\times$1080 \\
\em{SH2} & Shooter 2&  &  & & 800$\times$600 \\
\hline
\em{Sanc} & Sanctuary VR \cite{Sanctuary} & Steam &  Unity & DX11 & 1280$\times$800 \\
\hline
\em{AS} & Altspace VR \cite{Altspace} & Steam &  Unity & DX11 & 1920$\times$1080 \\
\hline
\em{OR} & VRMark \cite{vrmark} & Steam &  Unity & DX11 & 1728$\times$972 \\
&  Orange Room &&&& \\
\hline

\end{tabular}
}
\end{center}
\caption{VR benchmarks \label{table_1}}
\end{table}

\subsection{Experimental Setup}
\label{4b}
To show the inter-frame similarity (explained in Section \ref{sec:Background}) present in the VR applications, we run the applications on a desktop machine whose configuration is shown in Table \ref{table_3}. Our desktop has an Intel GPU. We instrumented the applications using \textit{Graphics Frame Analyzer}(GFA) \cite{Graphics}, a performance analysis tool provided by Intel. 
This tool helps us analyze every single frame in a stream of frames. In a graphics application to render a frame, the application issues several \textit{draw calls} which are nothing but function calls to the GPU. The GPU executes a pre-defined set of instructions for each draw call. We can think of this as a shader function. GFA profiles each frame of an application at the granularity of draw calls. Using this profiler, we collect the following information: contents of the rendered frame, number of draw calls required for rendering the frame, the shader/kernel executed by the GPU (instruction traces), and the number of instances of each draw call, i.e., software threads. Herein, lies a problem. The VR applications are written in Direct X 11, and thus we had to use an Intel core based desktop machine along with GFA to get the contents of all the rendered frames and additional information. The issue is that our simulator framework or other popular ones~\cite{bakhoda2009analyzing, power2014gem5} are designed for NVIDIA GPUs and can run CUDA programs. We thus modeled the Intel GPU architecture on our simulation setup as far as possible (same number of cores, etc.) and wrote a tool to translate Intel instructions to PTX instructions. Only a few instructions posed a problem in terms of their semantics; better explanations were found in other Intel manuals. With regards to the architecture, the NVIDIA Kepler architecture was found to be the closest (refer to Table \ref{table_3}).

Our simulator is a Java-based cycle-approximate GPGPU simulator, \textit{GPUTejas} \cite{gputejas}, which has been rigorously validated with native hardware. To get the latency of ATW, we implemented the ATW algorithm in CUDA and simulated it on GPUTejas. The details of the simulated scenarios are explained in Section \ref{3e}.

\begin{table}[]
\footnotesize
\begin{center}
\begin{tabular}{ c c} 

\hline
\textbf{Parameter} &
\textbf{Type/Value} \\ 
\hline\hline
\rowcolor[HTML]{EFEFEF}\multicolumn{2}{c}{Desktop Configuration} \\ \hline
CPU & 10\textsuperscript{th} Gen Intel\textregistered Core i7-10700 \\
GPU &  Intel\textregistered UHD Graphics 630 \\
VR Headset & Oculus Quest 2\\
\hline
\rowcolor[HTML]{EFEFEF}\multicolumn{2}{c}{Simulator Configuration} \\ \hline
GPU Architecture & Kepler \\
Clock Frequency  & 1 GHz \\
\#TPCs & 8 \\
\#SMs per TPC & 2\\
\#SPs per SM & 8\\
Warp Size & 32 \\
Shared memory size per SM & 16  KB\\
\hline
\end{tabular}
\end{center}
\caption{Details of the baseline system}
\label{table_3}
\end{table}

\subsection{Time Warping}
\subsubsection{Latency of the Time Warping Operation:}
\begin{figure}[]
    \centering
    \includegraphics[width=\columnwidth]{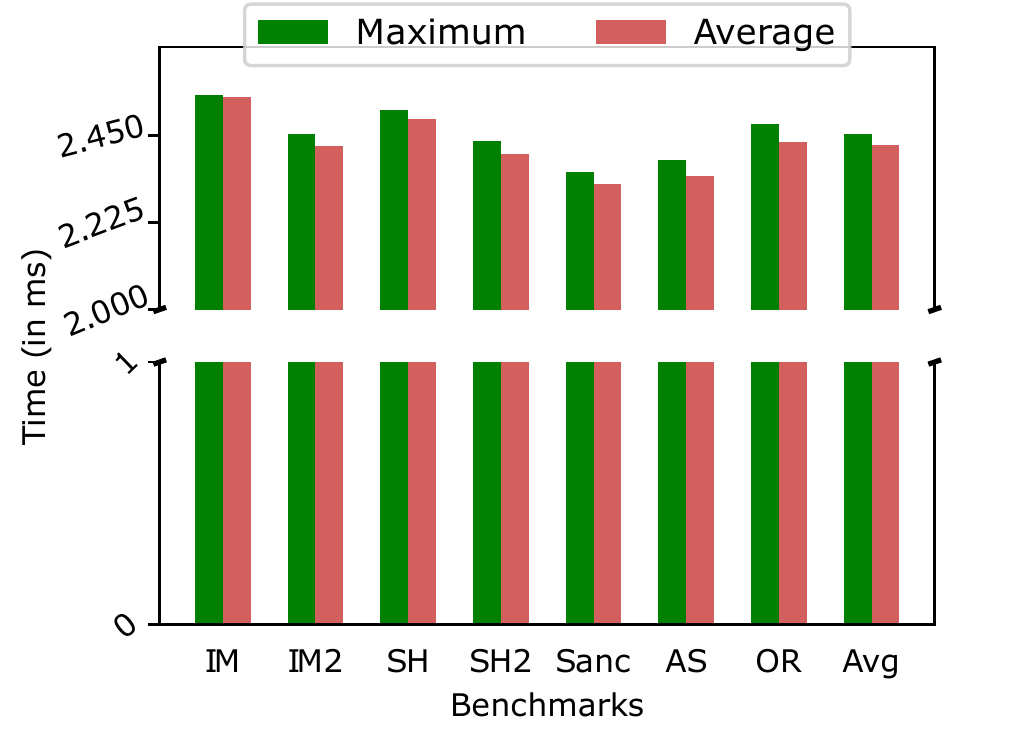}
    \caption{Latency of the time warping operation}
\label{fig_5}
\end{figure}

We wrote a CUDA program for the time warping function \cite{pim}. The inputs to this program are the frame that needs to be warped and the HMD (head mounted display) matrices at the render and display times. These HMD matrices are 4$\times$4 homogeneous matrices that contain the head's position at a given time. Then we run this kernel on the simulator for the frames collected by the GFA profiler and various head poses. In this case, we do not take consecutive frames rather we select those frames that are graphically more complex so that we have more variation in the inputs because consecutive frames are generally similar in nature as discussed in Section \ref{sec:Background}. Figure \ref{fig_5} shows that the execution time of the time warping kernel when all the resources are allotted to it.
 
We make the following observations from the figure:\\
\circled{1} For a particular benchmark, the maximum and average values are roughly the same, which means that the time warping kernel takes almost the same amount of time irrespective of the frame complexity and head poses. The maximum difference between the average and maximum value of the latency is 0.035 ms for all the benchmarks. \\
\circled{2} The average latency across all the benchmarks ranges from 2.32 ms to 2.55 ms; this  shows that the latency is almost independent of the frames' properties.
 
\subsubsection{Latency of GPU-Accelerated Asynchronous Time Warp (ATW):}
\label{3e}
\begin{figure}[]
    \centering
    \includegraphics[width=\columnwidth]{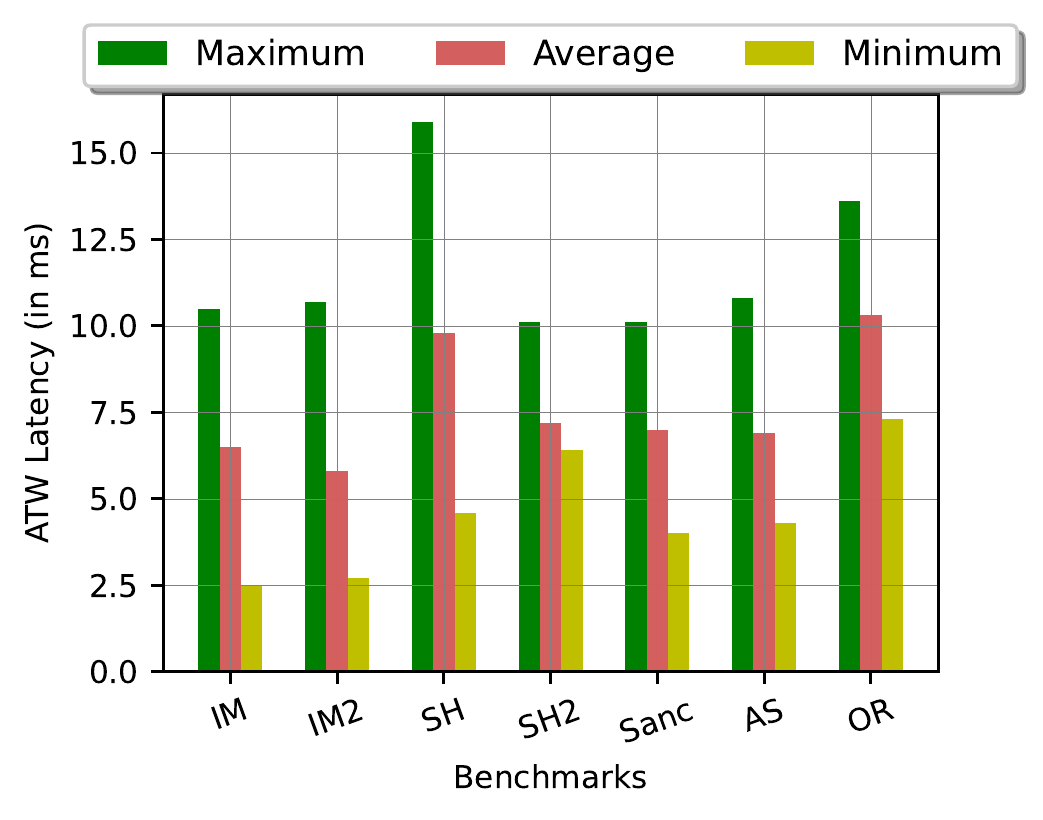}
    \caption{Variation in the ATW latency}
\label{fig_10}
\end{figure}

To evaluate the latency of ATW, we simulate two types of environments in GPUTejas: \circled{1}  All the cores are utilized by a single VR application which uses GPU-accelerated ATW. \circled{2} That VR application shares the graphics resources with another application. We do not consider the scenario where more than two applications run on the same GPU because it is not generally the case. Figure \ref{fig_10} shows the ATW latency of all the benchmarks. The large variation in the ATW latency for a particular benchmark and across the benchmarks shows that the ATW latency depends upon the frame whose rendering is getting preempted and the graphics resources' availability.

We make the following observations from the figure:\\
\circled{1} The maximum latency is almost 15 ms which is much larger than the latency of the time warping kernel ($\approx$ 2.5 ms) shown in Figure \ref{fig_5}. \\
\circled{2} There is a lot of difference among the minimum, maximum and average values of the latency for a benchmark, which means that the ATW latency varies a lot across frames and it is highly unpredictable and uncertain. The ATW latency ranges from 2 ms to almost 16 ms across benchmarks.

\subsection{Inter-Frame Similarity}
\label{4c}

To show the similarity between frames, we extracted 100 consecutive frames from every application.

\begin{figure}[!htb]
	\centering
	\includegraphics[width=0.9\columnwidth]{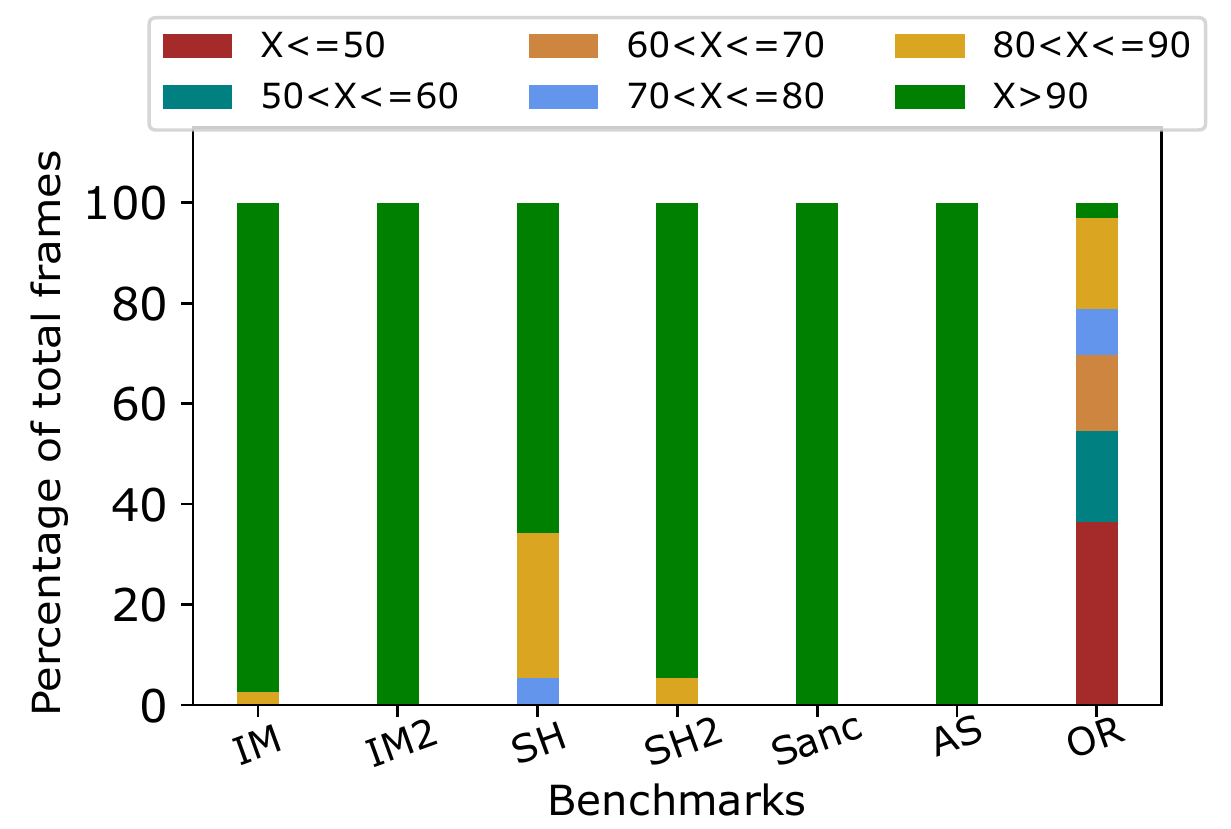}
	\caption{Macroblock similarity percentage between two consecutive frames (X)}
	\label{fig_6}
\end{figure}

\subsubsection{Frame-Level Similarity:} As mentioned in Section \ref{2b}, we use the macroblock similarity metric to find the similarity at the frame-level. Figure \ref{fig_6} shows the macroblock similarity percentage between  two consecutive frames. It is evident from the figure that for all the benchmarks except for one benchmark (\textit{OR}), all the frames are at least 70\% similar to their neighboring frames. Even for \textit{OR}, 60\% of the frames have at least 50\% inter-frame similarity.

\subsubsection{Feature-Level Similarity: } 
\begin{figure*}
\centering
\includegraphics[width=0.8\textwidth]{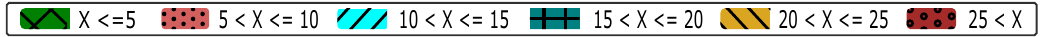}
\subfloat[InMindVR]{
  \includegraphics[width=0.33\textwidth]{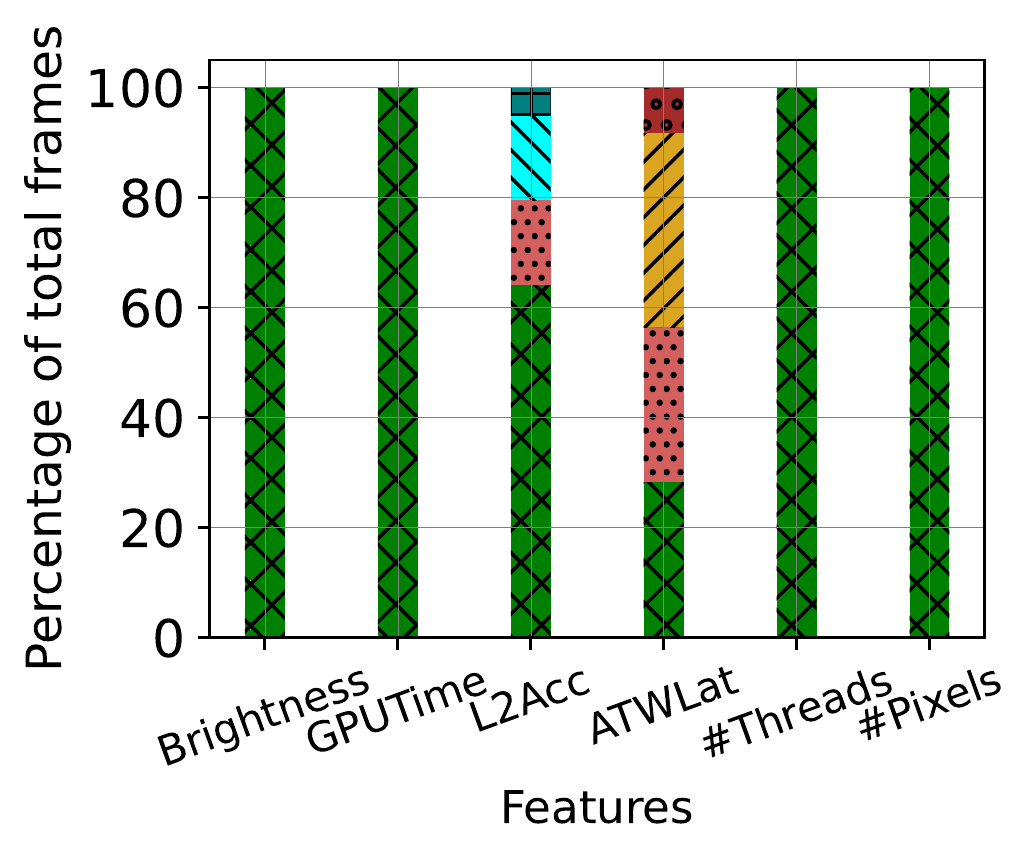}
}
\subfloat[InMindVR 2]{
  \includegraphics[width=0.33\textwidth]{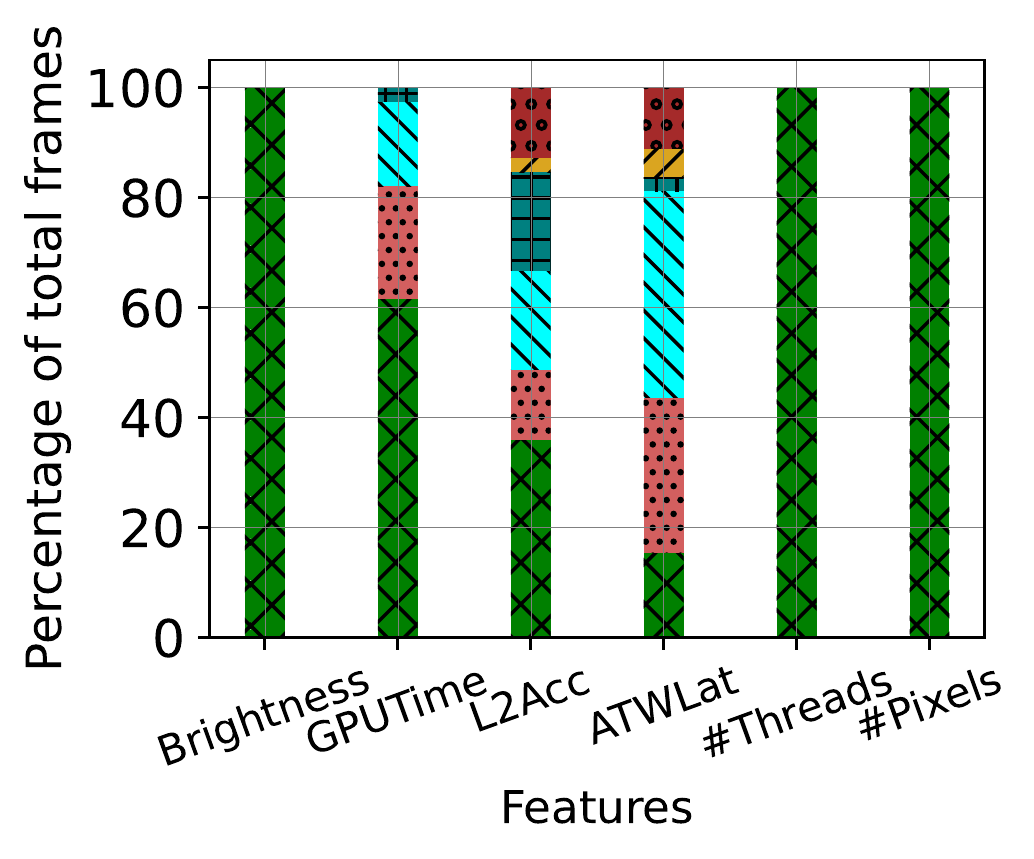}
}
\subfloat[Shooter]{
  \includegraphics[width=0.33\textwidth]{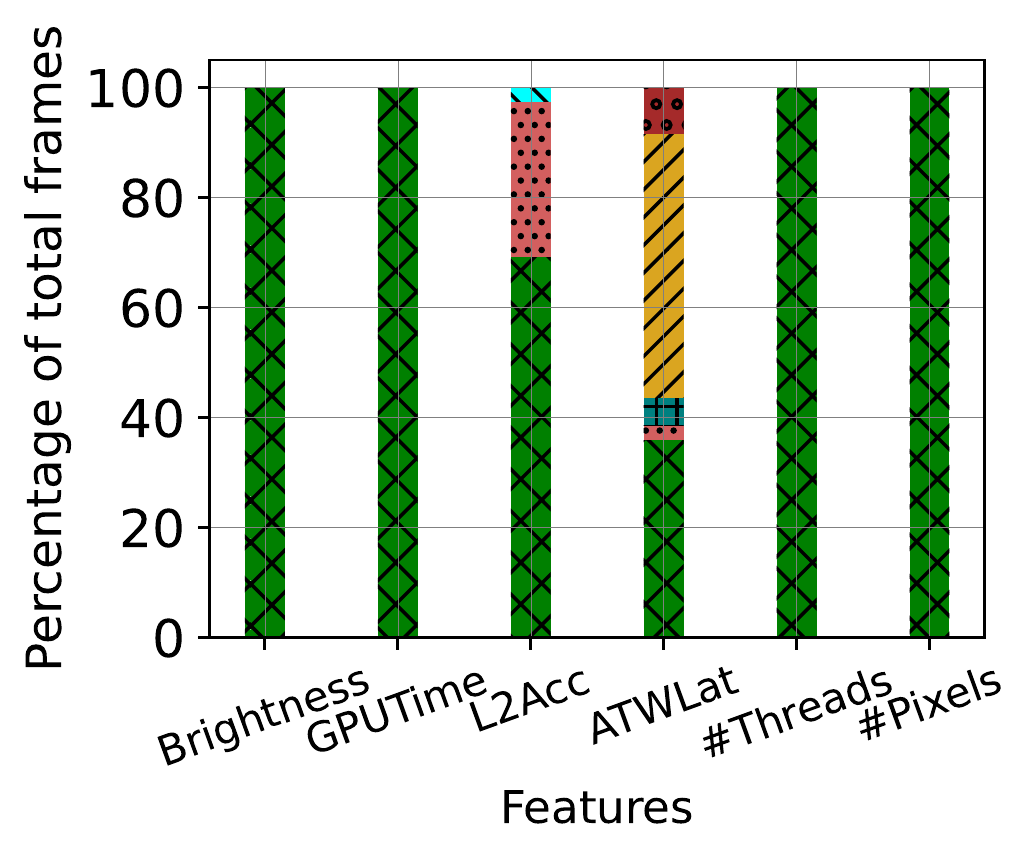}
}
\hspace{0mm}
\subfloat[SanctuaryVR]{
  \includegraphics[width=0.33\textwidth]{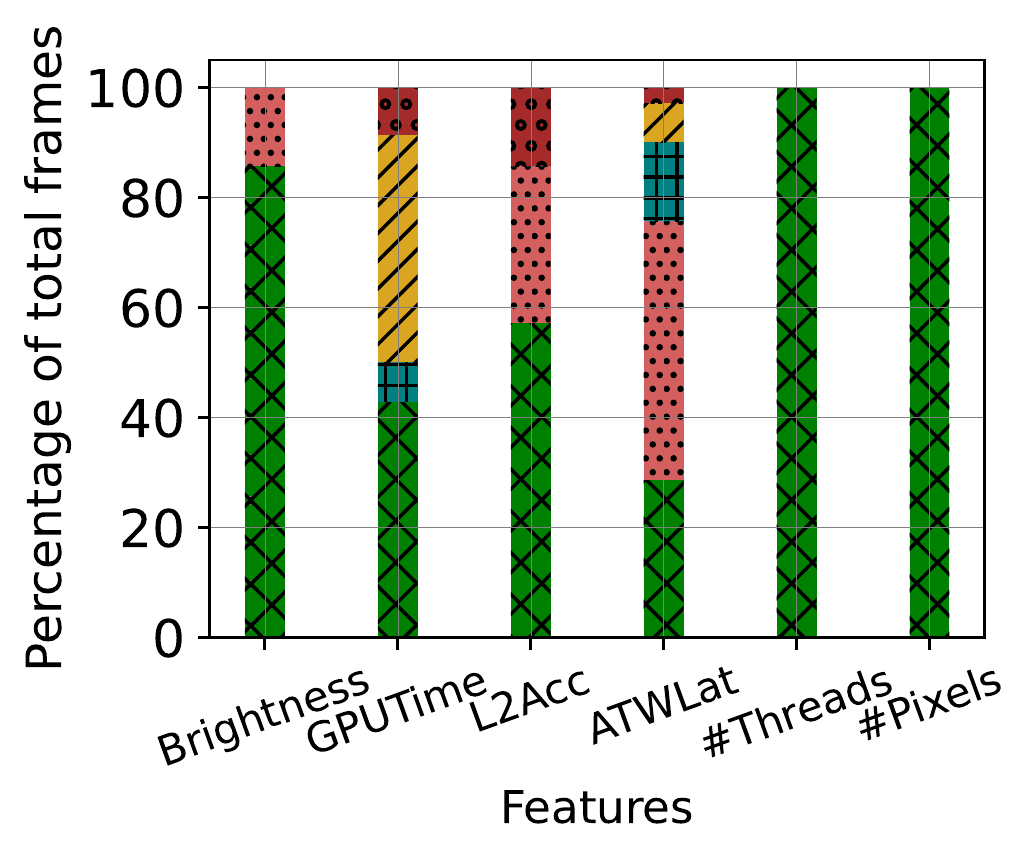}
}
\subfloat[AltSpaceVR]{
  \includegraphics[width=0.33\textwidth]{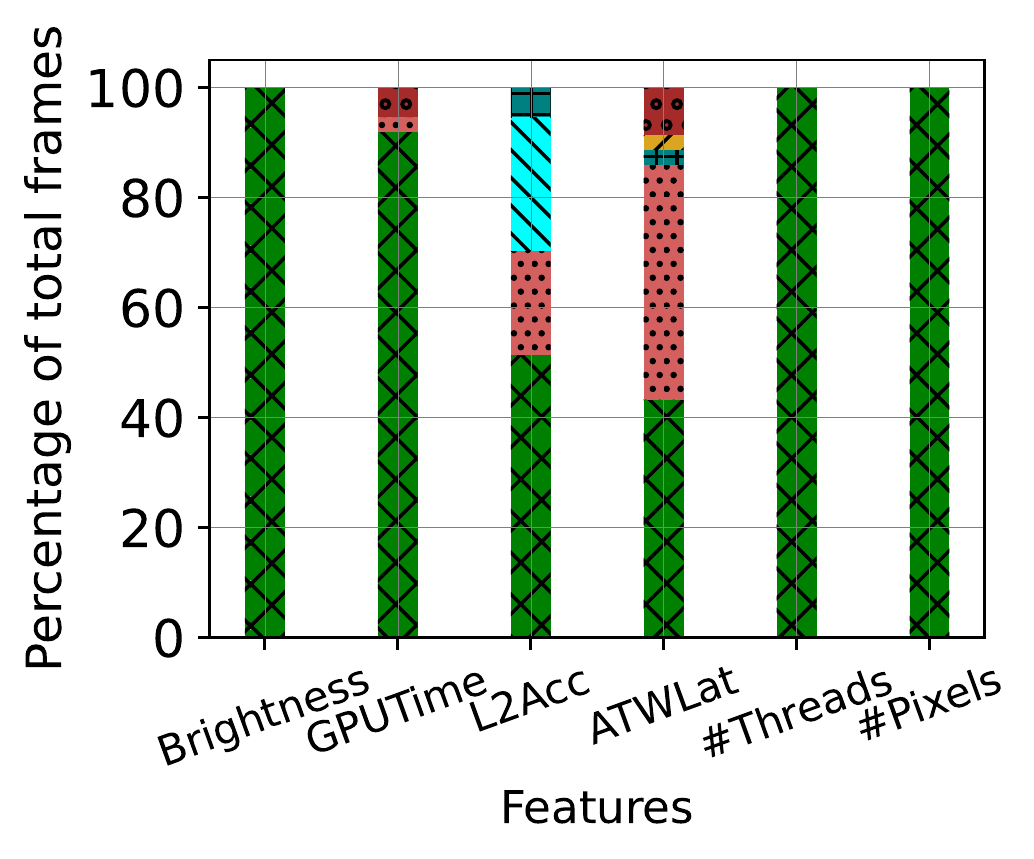}
}
\subfloat[VRMrak Orange Room]{
  \includegraphics[width=0.33\textwidth]{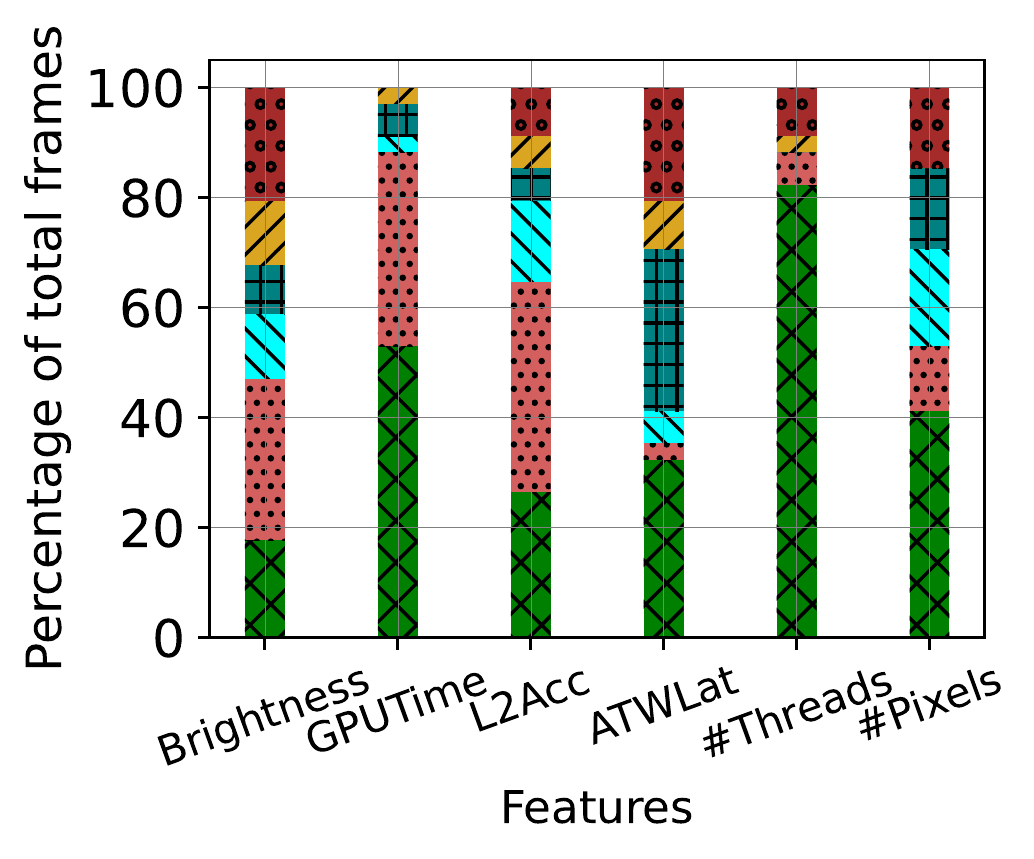}
}
\caption{Percentage variation from the last rendered frame's value (denoted by $X$)}
\label{fig_7}
\end{figure*}

To say whether two frames are similar or not, we collect a few frame-specific features/attributes and compare the corresponding values. These attributes are shown in Table \ref{table_4}. All these features are not extracted from the same source. For {\em brightness}, we take the rendered frame from GFA and calculate the value using the formula given in Ref. \cite{Python}; it is a function of the $RGB$ values. \#Threads and \#Pixels are collected directly from GFA. $GPUTime$, $L2acc$, and $ATWLat$ are obtained from the simulation results.  Figure \ref{fig_7} shows the similarity present in the neighboring frames of an application. 
Comparing the trends in the sub-figures, we make the following observations: \circled{1} For 80\% of the frames, variation in all the features' values from the previous frame's values is at most 25\%.
\circled{2}Features that show almost constant behavior across the frames are brightness, \#threads and, \#pixels.

\begin{table}[!htb]
\footnotesize
\begin{center}
	
\resizebox{0.99\columnwidth}{!}{	
\begin{tabular}{ |l|p{45mm}|l|} 

\hline
\rowcolor[HTML]{EFEFEF}
\textbf{Parameter} &
\textbf{Description} &
\textbf{Collected From} \\ 
\hline\hline
Brightness & Perceived brightness of the & GFA \\ & frame \cite{Python} &   \\
\hline
GPUTime & GPU execution time or & GPUTejas \\ & rendering time (in ms) &  \\
\hline
L2Acc & Number of L2 cache accesses while rendering the frame & GPUTejas \\
\hline
ATWLat & ATW Latency & GPUTejas \\
\hline
\#Threads & Number of software threads spawned for drawing the frame & GFA \\
\hline
\#Pixels&  Number of pixels rendered in a & GFA \\ & frame &  \\
\hline

\end{tabular}
}
\end{center}
\caption{Frame-specific attributes}
\label{table_4}
\end{table}

\subsection{Resolution of the Human Eye and Low MPD} 
\label{sec:lowmpd}

A human with 6/6 vision has a resolution of 0.59 arc minute per line pair~\cite{2018Notes} (resolution: $\approx$ 0.3 arc-minutes). Now assume that the head rotates at the rate of 200 degrees per second~\cite{sora} (in a fast VR game). The minimum eye resolution thus corresponds to 0.02 ms. If the MPD is more than that we will perceive a lag, which may lead to discomfort and visual disorientation. We have sadly experienced this on Oculus Quest 2 when we were playing a Harry Potter style game with a broom and an orb. This calculation is however theoretical and the vision system can compensate to some degree. Thus, the concept of JND (just-noticeable difference) has been proposed, which represents practical limits (measured in lab settings). The JND for regular yaw-based head motion has been measured to be as low as 1 ms~\cite{jerald-thesis} when there are other stimuli that last for more than 60 ms, and roughly 3.2 ms without other stimuli~\cite{3ms}. For haptics-VR applications for robotic arms, the JND limit is 2 ms~\cite{haptics}. Hence, this means that we cannot miss frame deadlines (one frame appears every 11 ms) or delay ATW by more than a few ms. The differences will be perceived by at least some users with good visual acuity. \textbf{This explains why in prior works \cite{pim}, the authors have tried to reduce the MPD by 4-5 ms and why the gaming industry is moving towards 144 to 360 Hz displays.}

\noindent
\\
\resizebox{0.9999\columnwidth}{!} {
\begin{tcolorbox}[colback=gray!10]
\parbox{\columnwidth}{\textbf{Insights:}\\
\circled{1}  The ATW latency varies from 2.5 ms to 16 ms across our benchmarks.  \\
\circled{2} There exists high inter-frame similarity in VR applications ($\approx$ 80\% of the time).\\
\circled{3} The human vision system is very sensitive to deadline misses and ATW delays beyond 3-5 ms.
}
\end{tcolorbox}
}

\section{Implementation}
\label{sec:Implementation}

As mentioned in Section \ref{sec:Background}, our aim is to predict the ATW latency in the case of a VR application. We broadly divide our implementation into various sub-components as follows: \circled{1} We first create an environment that simulates the GPU-accelerated ATW algorithm. \circled{2} We then identify the input features for the regression model.
\circled{3} We run the simulations to collect the sample data and divide the collected data into a training and testing set. \circled{4} We then train multiple predictors and compare their performance for the test data. \circled{5} After selecting the best performing predictor, at  run time we check its performance in terms of the refresh deadline miss rate and how close the completion of the time warping is to the display refresh point.

\subsection{Simulation Environment}
As mentioned in Section \ref{4c}, we use the \textit{GPUTejas} \cite{gputejas} simulator for our experiments. Table \ref{table_3} shows the simulator configuration. We  implemented two types of scenarios:  \circled{1} Only one benchmark runs with GPU-accelerated ATW on the simulator. Since, the display refresh rate is 90 Hz, in the period of every 11.11 ms, we need to preempt the frame's simulation and initiate the ATW operation. \circled{2} The main VR benchmark shares the cores with another application. For this, we divide the cores equally between the two applications and the cores are assigned to the running application if one application finishes.

\subsection{Identifying the Features}
\label{identify}
The important task in designing any prediction model is to define a set of input features that have a pattern which can be learned by the model and used for accurately predicting the output. Good feature selection is critical to the success of the model. In our case, the selected features must be correlated with the frames' characteristics and the amount of resource contention. The features should also have a correlation with the ATW latency that we want to predict.

In Section \ref{3e}, we observed that the ATW latency varies across frames due to the activities associated with the preemption of the current frame's rendering tasks, in other words on the current resource usage. Also, as shown in Figure~\ref{fig_10}, the variation in the ATW latency of  two consecutive frames is less that 20 \% for 80\% of the frames across all benchmarks. Hence, we can say that the ATW latency for the last rendered frame has a correlation with the ATW latency for the current frame. Also, it incorporates the effect of resource contention. So, we choose $PrevATWLat$ as one of the input features.

Similarly, the GPU execution time or rendering time of a frame depends upon the resources' availability and the frame's complexity (refer to Section \ref{3e}).  Hence, we make the GPU execution time an input feature. Since, the frame which the time warping preempts is not rendered completely, we do not have its execution time at the time of prediction. So, we take the execution time for the last rendered frame and this method works because as shown in Figure \ref{fig_10}, almost 90\% of the time, the variation in the execution time of two consecutive frames is less than 10\%

As mentioned earlier, the ATW latency depends upon the frame currently getting rendered and our model should be able to learn the frames' behavior. Hence, we consider a few more frame-specific metrices as input features. These features are \#vertices,\#pixels rendered, \#draw calls, and \#threads for those draw calls. We used these features to calculate the inter-frame similarity too in Section \ref{sec:Motivation}. To capture the memory access pattern while rendering the frame, we choose the number of L2 cache accesses as one of the features. Similarly, brightness is also a frame characteristic and we choose this also as one of the input features. One more reason to select this as a feature is that to set the brightness of a frame, a pixel shader is used that blends each pixel with predefined colour values. Depending upon how bright the frame is, the complexity of the shader varies and hence the preemption latency of the frame changes. Table \ref{table_6} shows all the input features. Note that we have a column that says whether the data is for the current frame or previous frame. Many features such as the rendering time or brightness can only be collected for the previous frame; they have not been computed for the current thread.

\begin{table}[]
\footnotesize
\begin{center}
\begin{tabular}{|l|p{34mm}|l|l|l|} 

\hline
\textbf{Feature} &
\textbf{Description} &
\textbf{Source}  & \textbf{Frame}
\\ 
\hline\hline

\textit{GPUTime} & Frame rendering time or &  GPUTejas & prev \\ 
\hline
\textit{L2Acc} & Number of L2 cache accesses &  GPUTejas & prev  \\ 
\hline
\textit{PrevATWLat} & ATW latency &  GPUTejas & prev  \\
\hline
\textit{\#Threads} & \# software threads &  GPUTejas & curr  \\
\hline
\textit{Brightness} & Perceived brightness of the frame\cite{Python} &  GFA & prev  \\    
\hline
\textit{\#Pixels} &  \# rendered pixels & GFA  & prev \\ 
\hline
\textit{\#Vertices} &  \# vertices in the frame & GFA & curr \\ 
\hline\textit{\#DrawCalls} &  Number of draw calls & GFA & curr \\ 
\hline

\end{tabular}
\end{center}
\caption{List of features}
\label{table_6}
\end{table}

\subsection{Prediction Model}

We implement four regression models, as explained in Section \ref{2c}. We use scikit-learn \cite{scikit}, an open-source machine learning library, for implementing these regression models.
The entire dataset is first divided into a training set and test set. These two sets are disjoint sets. The splitting of the dataset into the training and testing data is done randomly. In the training phase, the model does not have access to the test data. After training, the test data is fed to the model for checking its accuracy. Our results show that Decision Tree Regressor gives the least prediction error. Also, as mentioned in Section \ref{sec:Motivation}, the decision tree provides an explainable model and we require explainablity in our case. We check the importance of all the features in the prediction process and the sensitivity of the model with respect to some features. Finally, we wrote a CUDA kernel for the learned decision tree structure. 

\subsection{Spatial Multiplexing}
\label{5e}

To reduce the MPD, Smit et al. \cite{SmitLBF09,SmitLF10} propose a multi-process architecture in which the rendering process runs in parallel with the ATW process on the same GPU hardware. The resources are statically partitioned (spatial multiplexing). This avoids the need for preempting the rendering process. They implement this architecture on a single GPU by using NVIDIA's built-in \textit{spatial multiplexing} feature, MPS -- they divide the GPU's computational cores into rendering and ATW cores. 

We would like to criticize this approach.
A modern GPU is already fully used by rendering workloads; thus, dedicating cores for ATW incurs performance overheads and ends up reducing the rendering bandwidth available to VR applications \cite{pim}. Furthermore, ATW cores remain mostly unutilized; this represents a wastage of resources. Most GPU-based applications and runtimes are not that flexible (as of today) to dynamically resize themselves based on the number of available cores. 

Nevertheless, to show the effect of spatial multiplexing on the ATW latency we simulate it in our architecture. 

We consider three types of scenarios:\\
\circled{1} One eighth or 12.5 \% of the total cores are dedicated for ATW (\textit{SM1}).\\
\circled{2} One forth of the cores are assigned to ATW (\textit{SM2}). \\
\circled{3} Half of the cores are assigned to ATW (\textit{SM3}). 

We finally compare the ATW latency of these architectures with the GPU-accelerated ATW.

\section{Results and Analysis}
\label{sec:Evaluation}
Our proposed scheme that comprises predictors and  profiling code is implemented in software using CUDA libraries.
The reason for using a simulator is because we don't have a VR platform that will allow us to make such intrusive changes (similar to prior work~\cite{HPTS21}).

\subsection{Performance of Various Predictors}
\begin{figure}[!htb]
    \centering
    \includegraphics[width=\columnwidth]{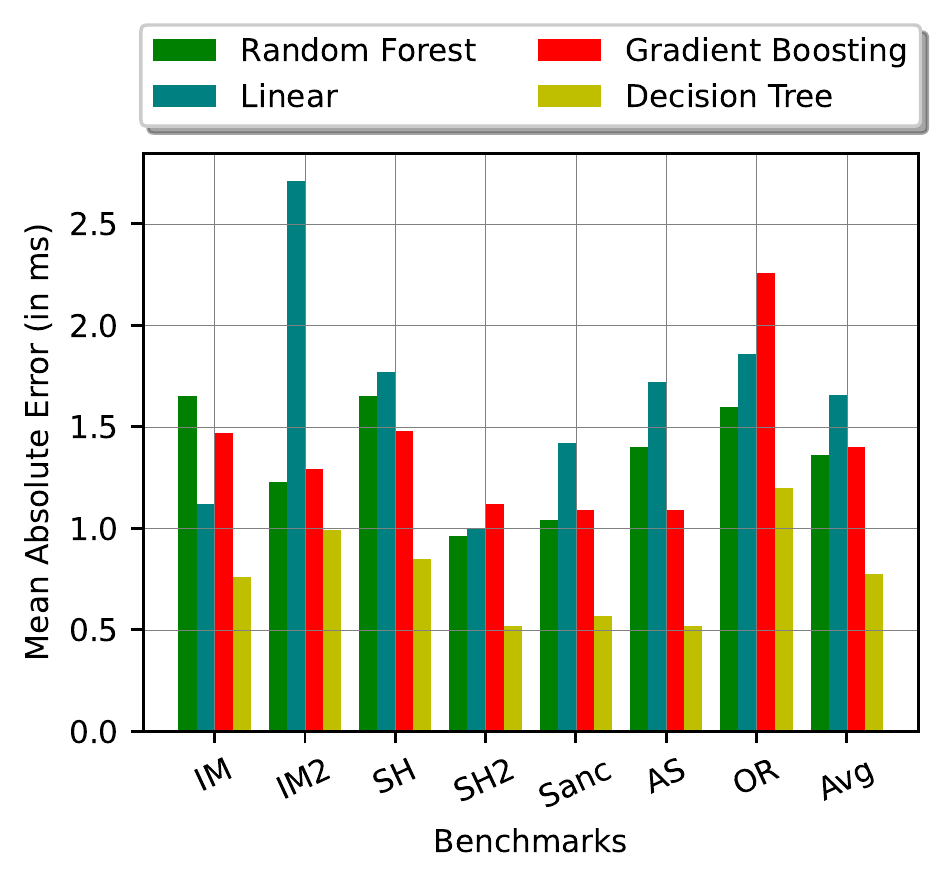}
    \caption{MAE of the different prediction models}
\label{fig_11}
\end{figure}

As mentioned in the previous section, we implement four prediction models which are random forest, linear regression, gradient boosting, and a decision tree. We use the same set of features for all the models. We compare the performance of these models in terms of the Mean Absolute Error (MAE). The formula for MAE is shown in Equation~\ref{eq2}. Figure~\ref{fig_11} shows the comparison of these four models on the basis of the mean absolute error in the prediction.  

\begin{equation} \label{eq2}
MAE =\frac{\sum_{i=1}^{N}|Y_i-\hat{Y_i}|}{N}
\end{equation}  

The results are shown in Figure~\ref{fig_11}. We observe that the decision tree-based regressor performs the best among all the models. It achieves an error of 0.77 ms on an average for all the benchmarks. 
The prediction error of the decision tree is less than 1 ms for all the benchmarks except for one benchmark, i.e., \textit{OR}. This is because its inter-frame similarity score is low.

Consider the case of the random forest predictor. The average error in this case is 1.4 ms, which is 81\% greater than the prediction error in the case of the decision tree. In spite of having 100 randomly created decision trees, we were not able to show a higher accuracy. This is because of the high inter-frame similarity and our features that reflected the same. There was no benefit to be gained by the randomness. On the other hand, each decision tree was trained with fewer samples, which secularly reduce the accuracy of all the constituent trees. 

Consider the case of linear regression that performs the worst for six out of seven benchmarks.  The average error we get for the linear regressor is 1.65 ms, 107\% greater than that of the decision tree.
We can infer that the relationship is far from linear. Hence, such simple models will not work.

Similar to random forest, the complex gradient boosting model has a prediction error that is 88\% more than the prediction error of the decision tree. The reasons are similar.

\subsection{Effect of the Predictor on the MPD}

To see the effect of our proposed approach on the MPD, we record
the deadline miss rate with and without the predictor. We also see how close the completion of the time warp 
operation is to the {\em display refresh point}. As mentioned in Section~\ref{sec:Implementation}, the display refresh rate is 90 Hz, which means that the display is refreshed every 11.11 ms. We analyze the refresh deadline miss rate for the following three cases:
\\ 
\circled{1} GPU-accelerated ATW that is the {\em Baseline} architecture. \\
\circled{2} ATW is invoked just after the frame is rendered or before 2.55 ms of the screen refresh operation, whichever occurs first. Recall that 2.55 ms is the average latency of the time warping kernel. (Refer to Figure~\ref{fig_5}) (we referred to it as the {\em Eager} approach) \\
\circled{3} ATW is invoked on the basis of the predicted ATW latency which is our proposed approach. We call it
{\em PredATW}. \\

\subsubsection{Refresh Deadline Miss Rate:}

\begin{figure}[]
    \centering
    \includegraphics[width=\columnwidth]{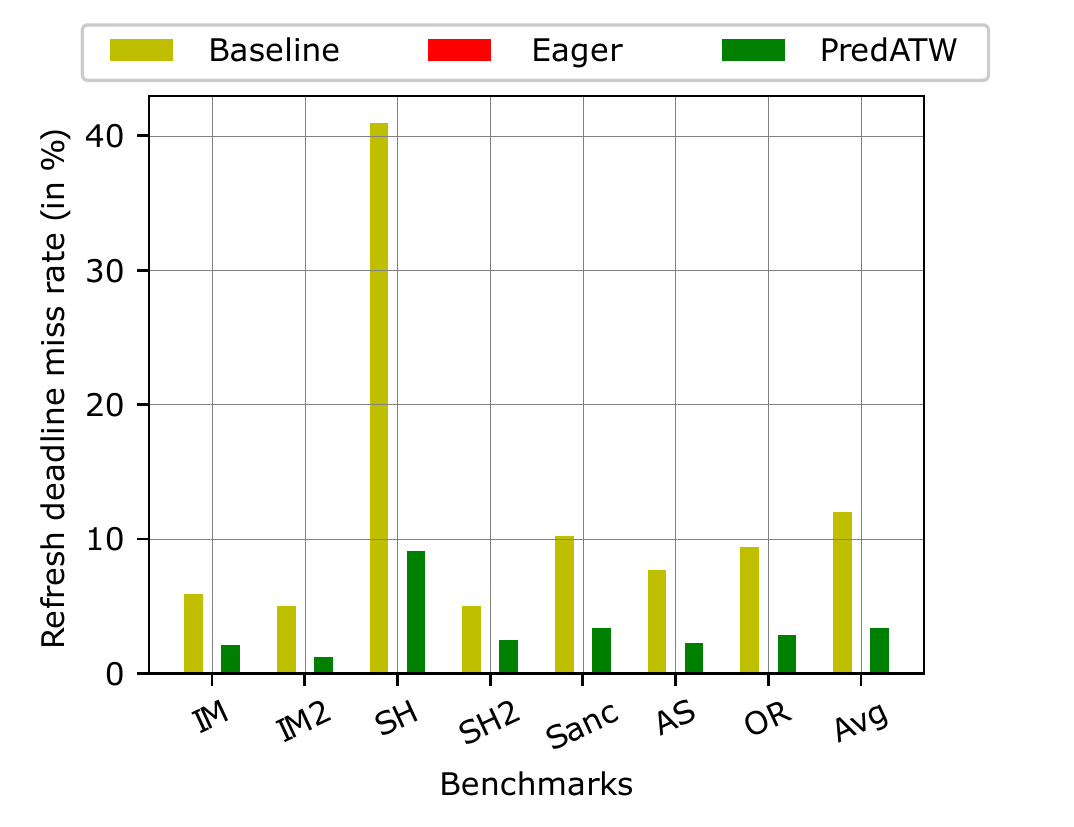}
    \caption{Effect of the predictor on the refresh deadline miss-rate. With {\em Eager}, we never miss any deadlines. Hence, it is at 0\% throughout.}
\label{fig_15}
\end{figure}

As mentioned in Section \ref{sec:Introduction}, when the ATW fails to get completed before the display refreshes, the MPD is increased and the user sees an anomalous or blurry view of the world. Figure \ref{fig_15} shows, the refresh deadline miss rates for the three cases described above. The figure shows that for all benchmarks, our predictor decreases the deadline miss rate. For GPU-accelerated ATW (the baseline), the deadline miss rate starts with 5\% and goes up to almost 41\% across all the benchmarks. The average miss-rate is around 12\%, which is significant.

For the {\em Eager} case where the ATW is invoked just after the frame is rendered, there are no deadline misses. However, as mentioned in Section~\ref{sec:Introduction}, this will ensure a large MPD and the reason for this is clear: we have a large gap between the time that ATW finishes and the display refresh point. The result shows that {\em PredATW} (our scheme) reduces the number of deadline misses substantially: from 12\% to 3.4\%. In this case, the miss rates are in the range of 2 to 4 \% for all the benchmarks except for one benchmark {\em Shooter}. However, if we compare the relative reduction, it is substantial. For the baseline it was around 40\%, we brought it down to 9.1\%. The overall reduction in the miss rate of {\em PredATW} is 73.1\%.

\subsubsection{ATW Completion $\rightarrow$ Display Refresh Point}

Let us now look at the average duration between the completion of ATW and the display refresh point:
the lower the better (subject to deadline misses).

\begin{footnotesize}
	\begin{center}
		\begin{tabular}{ |c|c|c|c|}
			
			\hline
			\textbf{Benchmarks} & {\bf Baseline} & {\bf Eager} & {\bf PredATW} \\
			\hline
			IM &  4.10  &  4.51 &  2.39\\
			\hline
			IM2 &  5.05  & 7.79 & 2.84 \\
			\hline
			SH &  1.45 &  2.93 &  2.53 \\
			\hline
			SH2 &  2.53 & 5.53 &   2.51 \\
			\hline
			Sanc &  3.40 & 7.49 & 1.55 \\
			\hline
			AS &  3.45 &  8.12 & 3.88 \\
			\hline
			OR & 2.58 &  4.68 &  3.15\\
			\hline
			\hline
			{\bf Mean} & 3.22 & 5.86 & 2.69 \\
			\hline
		\end{tabular}
	\end{center}
\end{footnotesize}

We observe that as compared to {\em Eager}, {\em PredATW} reduces this duration by 55\%. It
is also 17\% lower than {\em Baseline}. In some cases the latter is better; however, this is
often in an unsafe zone where deadlines are missed. The important point to note that for $PredATW$ this duration is on an average less than 3.88 ms all the time, which will most likely not be perceived (see Section~\ref{sec:lowmpd}). For the other two schemes, we saw a lot of high values (7-8 ms) in our results, which will cause deadline misses and even otherwise be perceived.

\subsection{Sensitivity Analysis }

As mentioned in Section \ref{4b}, we identify two features \textit{GPUTime} and \textit{PrevATWLat} to be the features that show the contention effect on the ATW latency. To show how these features affect the prediction error, we train and test the model with and without these features. 

\subsubsection{Effect of the GPUTime Feature:}
Figure~\ref{fig_13} shows the effect of the GPUTime feature on the prediction error (measured in terms of MAE). It can be observed from the figure that for all benchmarks, the prediction error decreases with the introduction of the GPUTime feature by roughly 33\%.  The trends are more or less the same across benchmarks.

\subsubsection{Effect of the PrevATWLat Feature:}
\begin{figure}[]
    \centering
    \includegraphics[width=\columnwidth]{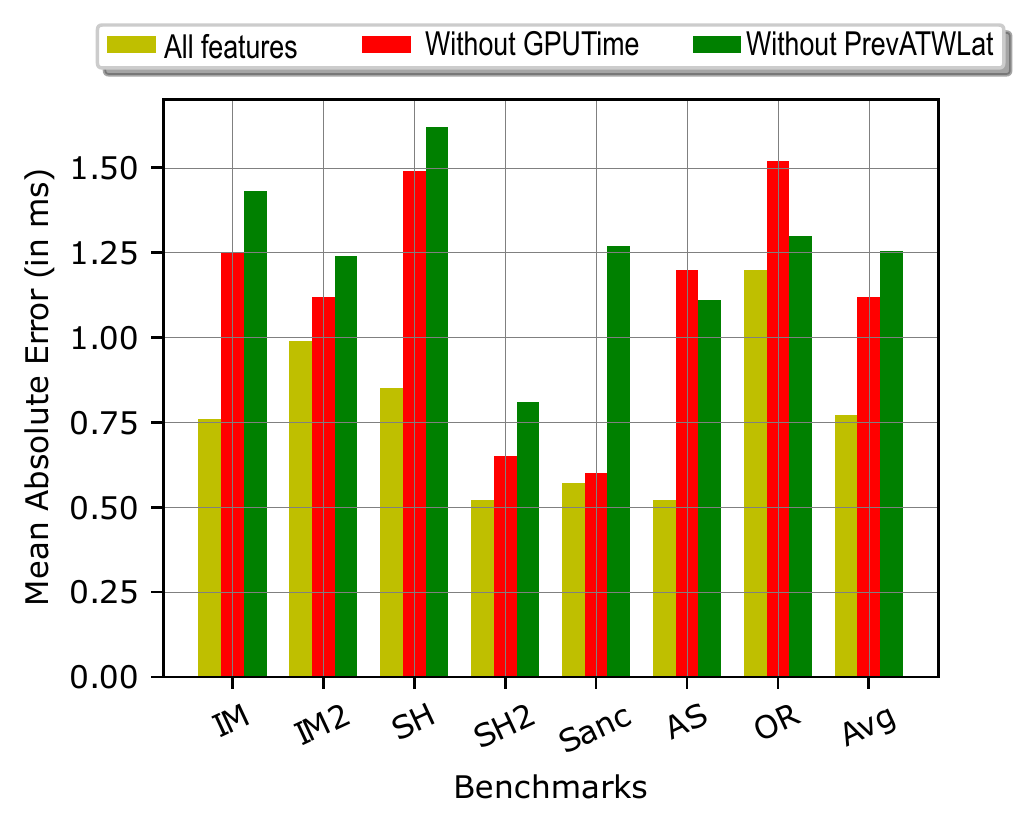}
    \caption{Effect of \textit{GPUTime} and \textit{PrevATWLat} on the prediction error}
\label{fig_13}
\end{figure}

The third bar in Figure \ref{fig_13} shows the effect of the PrevATWLat feature on the prediction error. Similar to  GPUTime, including PrevATWLat also has a positive effect on the prediction error. We observe that when we remove PrevATWLat from the list of input features, the prediction error increases by 62\%.

\subsubsection{Importance of Each Feature:}

\begin{figure}[]
    \centering
    \includegraphics[width=\columnwidth]{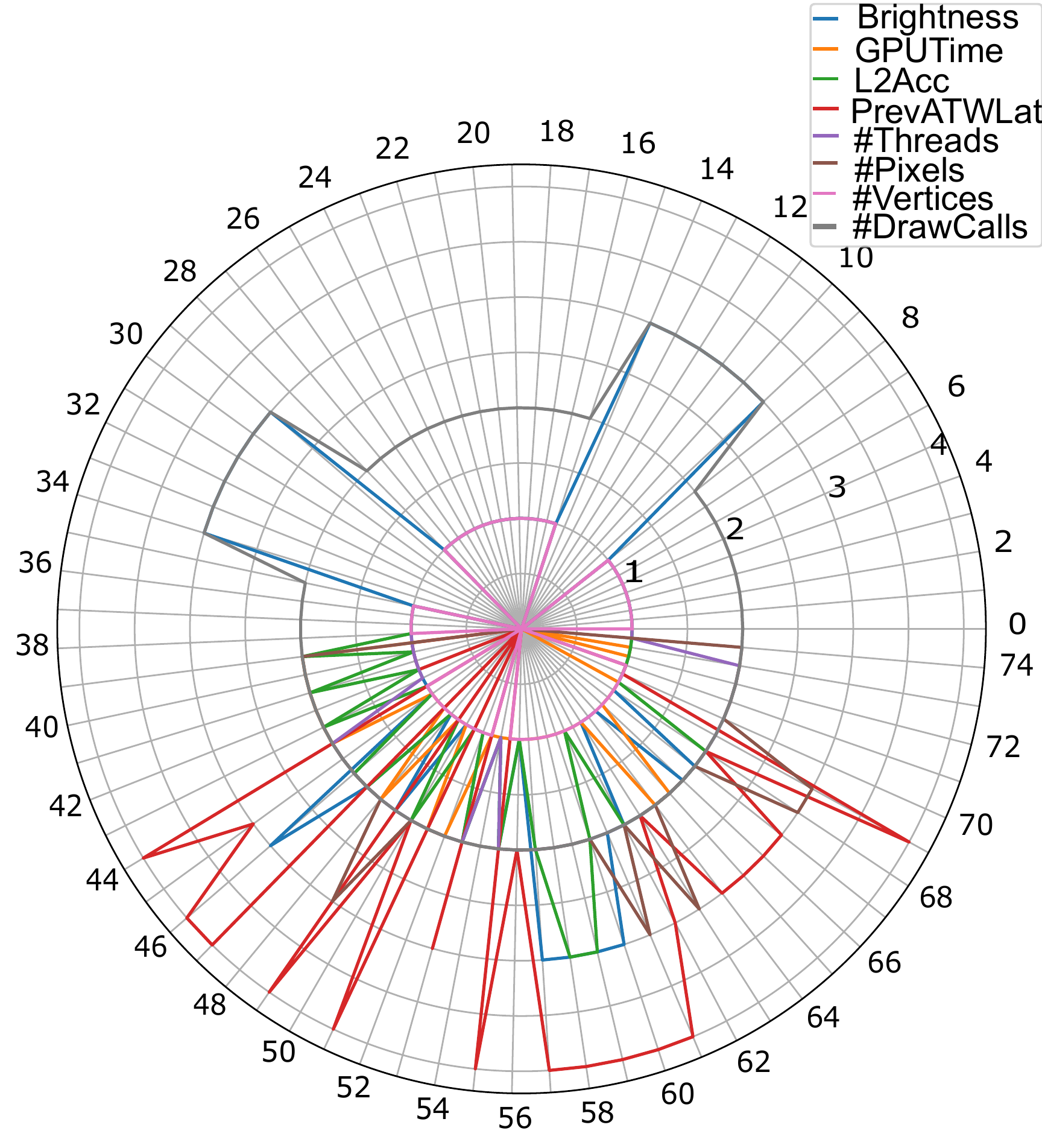}
    \caption{Frequency of each feature in the decision path for different test points}
\label{fig:radar}
\end{figure}

To show the importance of each feature in the prediction process, we find the frequency of each feature in the decision path for 75 representative test samples and plot them using the radar plot in Figure \ref{fig:radar}. The radar plot in the figure shows that three features -- \textit{Brightness} \textit{PrevATWLat} and \textit{\#DrawCalls} -- are used the maximum number of times in the decision paths. 

There are some interesting trends here. For the first 50\% of the benchmarks (0-37), only 2-3 features are used: Brightness, \#DrawCalls and \#Vertices. That too, they are used 1-3 times. This basically means that is easy to predict the ATW, and a quick decision can be made on the basis of these features. For the rest of the samples ($\approx 50$\%), almost all the features are used. 
Here, GPUTime plays a major role in the sense that it is close to the root and is used to make the first few decisions. PrevATWLat is at the other end of the spectrum -- it is used 2-4 times. This basically means that the predicted value is very sensitive to this parameter. The rest of the features have a moderate frequency and thus have a moderate amount of sensitivity.

\subsection{GPU-accelerated ATW vs Spatial Multiplexing}
\begin{figure}[!htb]
    \centering
    \includegraphics[width=0.8\columnwidth]{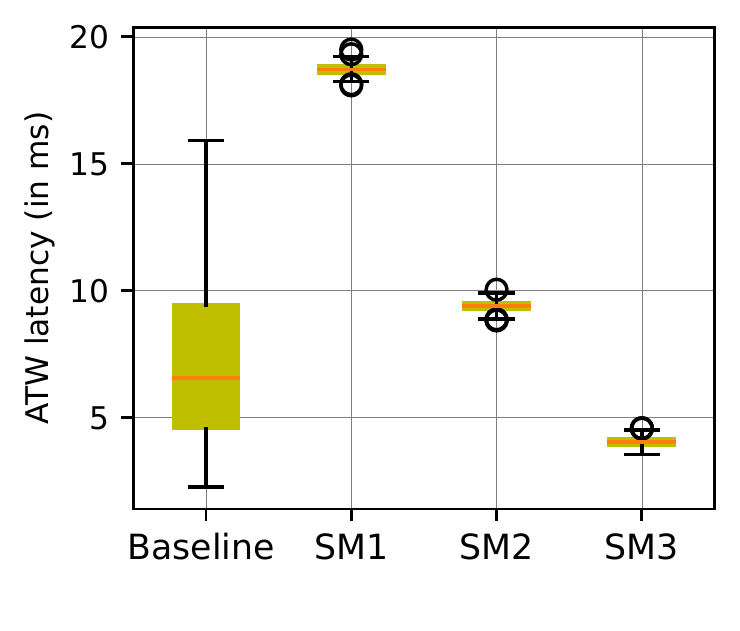}
    \caption{ATW latency for the baseline system and {\em SM1} (12.5\% cores for ATW), {\em SM2} (25\% cores for ATW), and {\em SM3} (50\% cores for ATW)}
\label{fig_14}
\end{figure}

We mentioned three forms of spatial multiplexing in Section \ref{5e}. Figure \ref{fig_14} shows the ATW latencies of those three cases  and the baseline configuration. It can be observed form the figure that the ATW latency varies from 2.5 ms to almost 16 ms which was also mentioned in Section \ref{3e} for the baseline system. Because of dedicated resources, the ATW variation in the three spatially multiplexed cases({\em SM1, SM2, and SM3}) is small. 

For {\em SM1} (12.5\% cores for ATW), the ATW latency is 18.71, which is way more than our frame time, 11.1 ms.
When we move to {\em SM2} (25\% reservation), the latency reduces to 9.39 ms, which is also very high.
The only figure that is borderline acceptable is for {\em SM3}, where the ATW latency is 4 ms.

\subsection{Latency of The Predictor}
\begin{figure}[!htb]
    \centering
    \includegraphics[width=0.76\columnwidth]{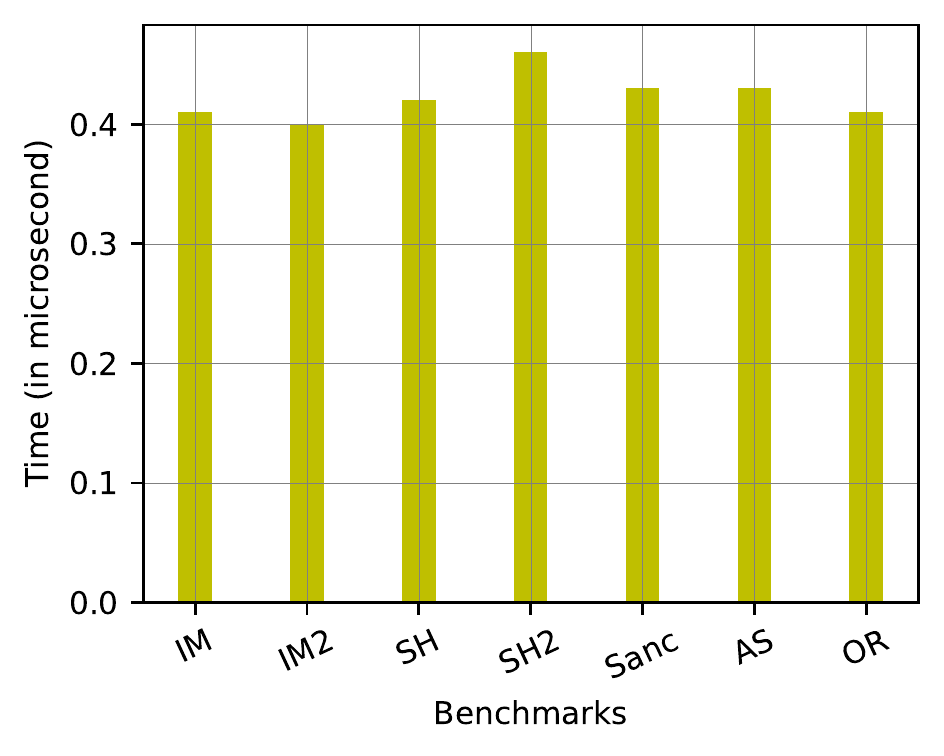}
    \caption{Maximum latency of the predictor}
\label{pred_lat}
\end{figure}

As mentioned earlier, we implement the learned decision tree structure in CUDA and run it on the simulator to evaluate the latency of the predictor. Figure \ref{pred_lat} shows the maximum latency of the prediction process for our benchmarks. It is clear from the figure that the predictor latency is a fraction of a microsecond. The latency is almost constant for all the benchmarks. The maximum value of the latency is 0.46 microseconds. Hence, we see that the predictor is not adding a significant delay to ATW.

\begin{table*}[!htb]
	\footnotesize
	\begin{center}

		\begin{tabular}{ | m{9em} | m{5em}| m{7em} | m{6em} | m{5em} |  m{5em} | m{5em}|} 
			
			\hline
			\rowcolor[HTML]{EFEFEF}
			
			\textbf{Work} &
			\textbf{ML-based Prediction} &
			\textbf{Predicted Data} &
			\textbf{Asynchronous Time Warp} &  
			\textbf{Outsourcing Time Warp} & 
			\textbf{Optimizing Time Warp} &
			\textbf{Optimizing Rendering} \\ [0.5ex] 
			\hline\hline
			
			Predictive Tracking \cite{KunduRP21, HouZBD19} & \checkmark & Head and body motions & $\times$ & - & - & $\times$ \\
			\hline
			
			FlashBack\cite{BoosCC16} & $\times$ & All possible viewing positions & $\times$ & - & - & $\times$  \\
			\hline
			
			Motion Predictions  + PDL \cite{SmitLBF09, SmitLF10} & \checkmark & Head and body motions & \checkmark &  \checkmark & \checkmark &  $\times$   \\
			\hline
					
			PIM-VR \cite{pim} & $\times$ &  $\times$ & \checkmark &  \checkmark & \checkmark &\checkmark  \\
			\hline
			Q-VR \cite{HPTS21}& $\times$ & $\times$ & \checkmark &  \checkmark & $\times$ &  \checkmark  \\
			\hline
			
			Schemes in \cite{PatneyKSKWBLL16,Luebke16} & $\times$ & $\times$  & $\times$ & - &  - & \checkmark  \\
			\hline
			
			\textbf{PredATW} & \checkmark & \textbf{ATW latency} & \textbf{\checkmark}  & \textbf{$\times$} & \textbf{$\times$}& \textbf{$\times$} \\
			\hline
		
		\end{tabular}

	\end{center}
	\caption{A comparison of related work}
	\label{table_2}
\end{table*}

\section{Related Work}
\label{sec:RelatedWork}

Since the MPD in VR systems adversely affects the user's experience, over the last few years, 
multiple approaches have been explored to reduce this latency for interactive VR systems. Recent works focus on  
\circled{1} 
predictive tracking \cite{KunduRP21, HouZBD19};
\circled{2} 
motion prediction + a programmable display layer (PDL) \cite{SmitLBF09, SmitLF10};
\circled{3} 
accelerating the time warp operation by offloading it to a separate hardware unit \cite{pim, HPTS21}; and
\circled{4} 
optimizing the rendering process \cite{PatneyKSKWBLL16, Luebke16}.
We present a brief comparison of related work in 
Table \ref{table_2}.

Recent works \cite{KunduRP21, HouZBD19} focus on machine learning-based approaches to anticipate viewing directions and positions to satisfy the ultra-low latency requirement of VR systems. Xueshi et al. \cite{HouZBD19} propose to do  predictive pre-rendering on the edge device. Kunda et al. \cite{KunduRP21} investigate a few common prediction algorithms, dead reckoning and kalman filtering, to predict future viewing directions and locations based on past behavior. FlashBack \cite{BoosCC16} predicts all the possible positions as well as orientations and pre-renders all views corresponding to these predictions. But pre-computing all the possible views and storing them in the memory increases the compute and storage overhead substantially. 

In this work, we use the latest pose obtained from the tracker device to warp the frames instead of predicting the head motion, which can be reasonably inaccurate. The only thing which we predict is the latency of ATW so that we can invoke the time warping kernel on time such that it finishes just before the display refresh point.

 Some works \cite{SmitLBF09,SmitLF10}  implement a dual-GPU architecture to avoid preemption. In this architecture, the first GPU, \textit{client}, renders application frames and generates motion information and the attributes of the pixels: colour, depth, and intensity. The data is sent over the PCIe bus to a shared system memory, and the second GPU, \textit{server}, uses this data to update the display device with a new frame. The second GPU, \textit{server} runs a custom program that takes the most recent application frame and data saved in shared memory and warps the frame based on the data and the current viewpoint. However, these solutions \cite{SmitLBF09,SmitLF10}  have high hardware overheads and also high power overheads because of the additional memory accesses and writes to the PCI bus.

One way to reduce the MPD significantly is to make the time warp operation fast. Waveren et al. \cite{Waveren16} investigate various hardware platforms for implementing a low-latency time warp algorithm. Xie et al. \cite{pim} propose PIM-VR, a Processing-In-Memory based ATW design that asynchronously executes ATW within a 3D-stacked memory without interrupting the rendering tasks on the host GPU. They also identify a redundancy reduction mechanism to further simplify and accelerate the ATW operation. Some works \cite{PatneyKSKWBLL16, Luebke16} optimize the rendering process itself to reduce the MPD. Toth et al. \cite{TothNA16} show that the number of rendered pixels can be reduced by up  to 36\% by rendering multiple optimized sub-projections without compromising on the visual quality. \textit{Foveation} is the quality degradation that increases with distance from the fovea (center of the retina) \cite{DBLP:journals/tog/GuenterFDTS12}. Foveated rendering algorithms exploit this phenomenon to improve performance \cite{PatneyKSKWBLL16, Luebke16}. These algorithms decrease rendering complexity and quality in the periphery while maintaining high image quality in the foveal part. Q-VR \cite{HPTS21} merges foveated rendering with the time warping operation to further improve the rendering performance.

\section{Conclusion}
\label{sec:Conclusion}

In the light of the pantheon of work in this field, we adopt a very different approach. Instead of focusing
on new rendering algorithms or predicting the head motion or using separate GPU hardware, we propose a 
relatively low-budget solution that requires only software support. We show that we can predict the ATW latency reasonably accurately in the presence of noise, even when another GPU application is running. The prediction error is limited to 0.77 ms, and the deadline miss rate is minimal ($\approx$ 3\%). Hence, our submission is that sophisticated solutions that require extra hardware support are not required. We would like to surmise that ATW latency prediction algorithms can further improve and with more data points or possibly with ensemble learning, the MAE error can be reduced to a value below 0.5 ms. This is a part of future work.

\bibliographystyle{IEEEtran}
\bibliography{references}

\end{document}